# Fully Two-Dimensional Incommensurate Charge Modulation on the Pd-Terminated Polar Surface of PdCoO$_2$

Pengfei Kong[1,*], Guowei Li[2,3,4,*], Zhongzheng Yang[1,*], Chenhaoping Wen[1,†], Yanpeng Qi[1,5,6], Claudia Felser[2], and Shichao Yan[1,5,†]

[1] School of Physical Science and Technology, ShanghaiTech University, Shanghai 201210, China

[2] Max Planck Institute for Chemical Physics of Solids, 01187 Dresden, Germany

[3] CAS Key Laboratory of Magnetic Materials and Devices, and Zhejiang Province Key Laboratory of Magnetic Materials and Application Technology, Ningbo Institute of Materials Technology and Engineering, Chinese Academy of Sciences, Ningbo 315201, China

[4] University of Chinese Academy of Sciences, 19 A Yuquan Rd, Shijingshan District, Beijing 100049, China

[5] ShanghaiTech Laboratory for Topological Physics, ShanghaiTech University, Shanghai 201210, China

[6] Shanghai Key Laboratory of High-resolution Electron Microscopy, ShanghaiTech University, Shanghai 201210, China

[*]These authors contributed equally to this work

[†]Email: wenchhp@shanghaitech.edu.cn; yanshch@shanghaitech.edu.cn

**ABSTRACT**

Here we use low-temperature scanning tunneling microscopy and spectroscopy to study the polar surfaces of $PdCoO_2$. On the $CoO_2$-terminated polar surface, we detect the quasiparticle interference pattern originating from the Rashba-like spin-split surface states. On the well-ordered Pd-terminated polar surface, we observe a regular lattice which has a larger lattice constant than the atomic lattice of $PdCoO_2$. In comparison with the shape of the hexagonal Fermi surface on the Pd-terminated surface, we identify this regular lattice as a fully two-dimensional incommensurate charge modulation that is driven by the Fermi surface nesting. More interestingly, we also find the moiré pattern induced by the interference between the two-dimensional incommensurate charge modulation in the Pd layer and its atomic lattice. Our results not only show a new charge modulation on the Pd surface of $PdCoO_2$, but also pave the way for fully understanding the novel electronic properties of this material.

**KEYWORDS:**
scanning tunneling microscopy, Pd-terminated polar surface, incommensurate charge modulation, Fermi surface nesting, moiré superlattice

Two-dimensional or quasi-two-dimensional system is one of the most fundamental platforms for the emergence of novel physical phenomena, including high-temperature superconductivity in layered cuprates and Dirac fermions in graphene.[1-3] The delafossite compound $PdCoO_2$ is another quasi-two-dimensional material with many peculiar electronic properties.[4-27] It consists of highly conductive Pd layers separated by the insulating $CoO_2$ layers, which results in large anisotropic electrical conductivity.[4, 8] At room temperature, the in-plane conductivity of $PdCoO_2$ is even higher than that of the alkali metals.[7,10,15] In $PdCoO_2$, a very dispersive Pd-derived band crosses the Fermi level (Figure 1b), resulting in a two-dimensional bulk Fermi surface as a nearly perfect hexagon.[8,17,27] The hexagonal Fermi surface in $PdCoO_2$ leads to fascinating quantum transport properties such as oscillatory magnetoresistance, hydrodynamic transport and super-geometric electron focusing.[8,10,14,19,22]

Cleaving the $PdCoO_2$ single crystals results in both the $CoO_2$- and Pd-terminated polar surfaces with the corresponding surface states. On the $CoO_2$-terminated polar surface, the surface states with Rashba-like spin splitting have been detected by the angle-resolved photoemission spectroscopy (ARPES)[16] and the quasiparticle interference (QPI) in scanning tunneling microscopy (STM) measurements.[26] On the Pd-terminated polar surface, a flat band has been detected in the ARPES measurements, and it has been suggested to be a spin-polarized band that induces surface ferromagnetism.[17] Although the Pd layer plays an important role in the electronic properties of $PdCoO_2$, the previous STM measurements only show an inhomogeneous Pd-terminated surface with unknown electronic structure[26] and the detailed microscopic information of the Pd-terminated surface is still missing.

Here we use low-temperature STM to systematically study the polar surfaces of $PdCoO_2$. Our STM data provide complete information for both the $CoO_2$- and Pd-terminated polar surfaces. On the $CoO_2$-terminated polar surface, we observe the QPI pattern resulting from the Rashba-like spin-split surface states. On the well-ordered Pd-terminated surface, we observe a Fermi-surface-nesting-induced two-dimensional incommensurate charge modulation (ICCM). This ICCM can interfere with the atomic lattice in the Pd layer, which forms a moiré pattern.

Figure 1c is the constant-current STM topography taken on a large area of $PdCoO_2$, which clearly shows two kinds of surfaces with different surface roughness. The lattice structure can be seen on the rough surface shown in Figure 1c. Although both the two surfaces show atomic triangular lattice with a few atomic defects (Figures 1d and 1e), the two layers have different step heights (as shown in the line profile across the atomic steps, inset of Figure 1c). The height of the rough-surface layer (~250 pm) is smaller than that of the smooth-surface layer (~330 pm). Since the $CoO_2$ layer consists of the edge-sharing $CoO_6$ octahedra and the Pd layer only contains a single-layer triangularly coordinated Pd atoms (Figure 1a), we would think the thick layer (~330 pm) with smooth surface is the $CoO_2$ layer and the thin layer (~250 pm) is the Pd layer (Figure 1c).

We next perform scanning tunneling spectroscopy (STS) to further confirm the surface terminations. For the $CoO_2$-terminated surface, the previous ARPES measurements have shown that the surface states in the $CoO_2$-terminated surface are split off by ~70 meV due to the inversion-symmetry breaking at the surface and the spin-orbit coupling.[17] In the $dI/dV$ maps taken on the smooth surface shown in Figure 1c, the electronic standing wave patterns of the surface states can be clearly seen (Figure 2a and Figure S1). The QPI pattern can be obtained by performing Fourier transform (FT) to the $dI/dV$ maps (inset of Figure 2a). By plotting a linecut along the Γ-K direction as a function of bias voltage, we observe a hole-like dispersion crossing the Fermi level with the band top at ~50 meV, which is consistent with the surface states of the $CoO_2$ surface reported in the previous STM experiment.[26] Because of the selected scattering rules, only one hexagonal ring-like feature appears in the QPI pattern. In the $dI/dV$ spectrum, the band top of the surface states in the $CoO_2$-terminated surface appear as a peak located at ~50 mV (Figure 2b). Our STS data clearly indicate that the smooth surface shown in Figure 1c is the $CoO_2$-terminated surface.

Having identified the $CoO_2$-terminated surface, we turn to the Pd-terminated surface that has larger surface roughness. At first glance, we would think that the regular lattice shown on the Pd-terminated surface (Figure 1e) is the atomic lattice of the Pd atoms, and it should have the same lattice constant as the atomic lattice of the $CoO_2$ surface (Figure 1d). However, as we can see from the line profiles taken on these two surfaces (Figures 2c and 2d), although the lattice constant of the $CoO_2$-terminated surface (~2.8 Å) is consistent with the atomic lattice of $PdCoO_2$, the Pd-terminated surface has larger lattice constant (~3.9 Å). The ratio between these two lattice constants is ~0.72. Furthermore, the corrugation of the regular lattice on the Pd-terminated surface is about ten times larger than that of the $CoO_2$-terminated surface (Figures 2c and 2d). Figures 2e-2h are the bias-voltage-dependent STM topographies taken on the same region of the Pd-terminated surface (the bias-voltage-dependent STM topographies on the $CoO_2$-terminated surface are shown in Figure S2). As the bias voltage is changed from +500 mV to +150 mV, some stripe-like patterns start to appear near the impurities (Figure 2f). In the STM topographies taken with the negative bias voltages (Figures 2g and 2h), a new superstructure with larger periodicity (~1.2 nm) emerges. From the STM topographies taken near the step edges, we can see that this kind of superstructure does not exist on the $CoO_2$-terminated surface (Figure S3).

In order to understand the regular lattice and the superstructure on the Pd-terminated surface, we perform FT to the STM topographies taken on the Pd-terminated surface to reveal their periodicities (see Figure S4 for more FT images). Figures 3a and 3b are the FT images of the STM topographies in Figures 2g and 2h, respectively. There are three sets of wave vectors marked by the yellow ($Q_0$), blue ($Q_1$) and green ($Q_{Bragg}$) circles. The $Q_{Bragg}$ wavevector corresponds to the atomic lattice of $PdCoO_2$ with a periodicity of ~2.8 Å. The $Q_0$ wavevector is along the $Q_{Bragg}$ direction, and the angle between the $Q_1$ and $Q_{Bragg}$ wavevectors is 30°. As shown in the line-cut profiles along the directions of the $Q_1$ and $Q_{Bragg}$ in the FT images of the STM topographies taken with various bias voltages (Figures 3c and 3d), the $Q_0$ and the $Q_1$ wavevectors are bias voltage independent. The ratio between $Q_0$ and $Q_{Bragg}$ is ~0.22 ± 0.02, and the ratio between $Q_1$ and $Q_{Bragg}$ is ~0.71 ± 0.02. This demonstrates that the $Q_1$ wavevector corresponds to the regular lattice of ~3.9 Å and the $Q_0$ wavevector is related to a real-space pattern with a periodicity of ~1.2 nm.

Our STM data in Figures 3a-3d shows that the $Q_1$ wavevector corresponds to the pattern with the periodicity of ~3.9 Å shown in the bottom panels of Figures 2c and 2d, and it is likely related with a charge modulation that is incommensurate with the atomic lattice. Fermi surface nesting is usually a good start point for exploring the origin of charge order. The previous ARPES measurements demonstrate that the Fermi surface topology of the Pd-terminated surface consists of the bulk Fermi surface and the smaller-size Fermi surface related with the surface bands.[17] Figure 3e shows the Brillouin zone and the measured Fermi surface (from Ref. 17) contributed by the Pd-terminated surface of $PdCoO_2$. In Figure 3e, the green arrow indicates the Bragg wavevector and the blue arrow shows the scattering wavevector between the parallel segments of the measured Fermi surface. The angle between the two wavevectors in Figure 3e is 30° and the ratio between them is ~0.66 ± 0.02 (Figure 3e). The Fermi surface nesting strength can also be reproduced in the autocorrelation of the Fermi surface shown in Figure 3e (Figure 3f). As shown in Figures 3c and 3e, the strongest nesting vector is consistent with the $Q_1$ wavevector shown in Figures 3a and 3b. The slight difference between the $Q_1$ wavevector in the STM topography (~0.71 ± 0.02) and the nesting vector in the measured Fermi surface (0.66 ± 0.02) could be because our $PdCoO_2$ sample has larger Fermi surface than that used in the previous ARPES measurements.

Since the $Q_1$-wavevector-related charge modulation is incommensurate with the atomic lattice, it creates an interference pattern with the atomic lattice and results in a moiré superlattice with larger periodicity. The $Q_0$ wavevector shown in Figures 3a and 3b corresponds to the moiré superlattice of the ICCM and the atomic lattice. This moiré effect can be confirmed by simply overlapping the atomic lattice and the ICCM lattice to deduce the superposition pattern shown in Figure 3g. Figure 3h is the FT image of the superposition pattern shown in Figure 3g. We can clearly see that Figure 3h matches very well with the FT images shown in Figures 3a and 3b, which proves that the $Q_0$ wavevector indeed results from the moiré superlattice of the ICCM and the atomic lattice. When performing variable temperature measurements, we find that this moiré superlattice persists even at 40 K (Figure S5). The existence of the moiré superlattice also indicates that the $Q_1$-wavevector-related regular pattern is not due to the surface reconstruction of the Pd atoms, because the periodicity of the surface reconstruction pattern is usually commensurate with the atomic lattice and it cannot further form the moiré superlattice. The $Q_0$ wavevector is almost invisible in the FT image taken with high bias voltages, such as 500 mV (Figure S6). This could be because the density of states from the Pd layer increases as increasing the bias voltage, which makes it difficult to detect the influence of the atomic lattice in the STM topographies taken with high bias voltages (Figure S6).

After understanding the ICCM and the moiré superlattice on the Pd-terminated surface, we next perform the d$I$/d$V$ measurements to reveal the electronic structure of this surface. As shown in Figure 4b, the two peak-like features located at −100 mV and +150 mV are likely related with the flat bands on the Pd-terminated surface (see Figure S7),[17] and there is still a finite density of states between the two peak-like features (see Figure S8 for the d$I$/d$V$ maps at the energies near the Fermi level). More interestingly, as shown in the bias-voltage dependent STM topographies (see Figures 2e-2h and Figure S9), the impurity-induced stripe-like patterns appears more prominent in the STM topographies taken with −100 mV and +150 mV, which indicates these stripe-like patterns are related with these flat bands on the Pd-terminated surface. In the previous theoretical calculations, in order to fit the ARPES data, the flat bands on the Pd-terminated surface are suggested to be the spin-polarized bands that are driven by the Stoner instability.[17] Since the STM tips used in this work are not spin-polarized, the structures shown in Figures 2e-2h should not be directly related with the magnetic properties of the Pd-terminated surface. Further spin-resolved STM/STS measurements under magnetic field and theoretical calculations including the ICCM are needed in the future to fully understand the origin of the flat bands on the Pd-terminated surface.

Our STM data indicate that the ICCM on the Pd-terminated surface is likely to be a two-dimensional incommensurate charge density wave driven by the Fermi surface nesting. However, no clear contrast inversion is seen in the bias-dependent topographies taken on the Pd-terminated surface (see Figure S4). This could be due to the coexistence of the ICCM, impurity-pinning-induced pattern and the moiré superlattice on the Pd-terminated surface, which makes it difficult for us to clearly identify the contrast inversion in the STM topographies. We expect more experimental observations in the future to provide stronger evidence to confirm the charge density wave nature of the ICCM on the Pd-terminated surface of $PdCoO_2$.

Finally, we also observe the inhomogeneous Pd-terminated surface that is similar as that reported in the previous STM work (Figure 4c).[26] In the STM topography on the same region as Figure 4c taken with the negative bias voltage (Figure 4d), the individual randomly distributed Pd atoms can be clearly seen. This indicates that the inhomogeneous Pd-terminated surface is an incomplete surface that results from the missing of the Pd atoms during the cleaving process. We also find that the d$I$/d$V$ spectra taken on this kind of Pd surface are spatially inhomogeneous (Figure S10).

In summary, we report a systematic study about the polar surfaces of $PdCoO_2$. We observe the QPI of the surface states in the $CoO_2$-terminated surface. On the ordered Pd-terminated surface, we find the ICCM that is driven by the Fermi surface nesting. This charge modulation pattern can interfere with the atomic lattice and induce the moiré superlattice. Since it has been proposed that incommensurate charge density wave can induce hydrodynamic transport in the clean electronic system,[28] the ICCM in the Pd layer may be helpful for understanding the hydrodynamic electron flow in $PdCoO_2$.[14] We expect the ICCM and the moiré superlattice would also exist on the Pd-terminated surface of $PdCrO_2$ which has similar Fermi surface as $PdCoO_2$.[29,30] Our work not only demonstrates a new charge modulation on the Pd-terminated surface of $PdCoO_2$, but also provides clues for further understanding the complex electronic properties in this class of delafossite metals.

Note: during the reversion, we became aware of a combined ARPES and STM study which shows the QPI on the flat Pd surface of $PdCoO_2$.[31] There are surface Pd vacancies on that Pd surface,

which may perform doping effect and influence the ICCM on the Pd surface. The more detailed comparison between the Pd surface reported in Ref. 31 and the Pd surface showing the ICCM needs further investigations.

## MATERIALS AND METHODS

**Synthesis of PdCoO$_2$:** Single crystals of PdCoO$_2$ were grown in an evacuated quartz ampule containing a mixture of high purity PdCl$_2$ (99.995 %) and CoO (99.99 %) through the following metathetical reaction: PdCl$_2$ + 2CoO $\rightarrow$ 2PdCoO$_2$ + CoCl$_2$. The ampule was heated to 930 °C in 2 hours and then to 980 °C in 4 hours. After a heat treatment of 0.1 hour, the tube was cooled down to 580 °C at a rate of 2 °C/hour. Finally, the system was heated to 700 °C again and stayed at this temperature for 3 or 4 days. The resultant crystals were washed with hot ethanol to remove CoCl$_2$.

**STM/STS measurements:** STM experiments were performed with a Unisoku 1200 low-temperature STM. PdCoO$_2$ single crystals were cleaved at 77 K with the pressure under $2 \times 10^{-10}$ Torr, and then immediately transferred into the STM for measurements at 4.3 K. Chemically etched tungsten tips were used for the measurements. STS measurements were done by using standard lock-in technique with a 3 mV modulation at the frequency of 914 Hz.

## ASSOCIATED CONTENTS

**Supporting Information:**

d$I$/d$V$ maps and their FT images taken on the CoO$_2$-terminated surface; Bias-dependent STM topographies taken on the CoO$_2$-terminated surface; STM topographies taken at the step edges; FT images of the STM topographies taken on the Pd-terminated surface; Temperature-dependent measurements up to 40 K; FT images for STM topographies taken with high and low bias voltages; Comparison with the ARPES data measured on the Pd-terminated surface; d$I$/d$V$ maps and their FT images taken on the Pd-terminated surface; Bias-dependent STM topographies taken on the Pd-terminated surface; d$I$/d$V$ spectra taken on the inhomogeneous Pd-terminated surface.

## AUTHOR INFORMATION

**Corresponding authors**
Correspondence and requests for materials should be addressed to C.W. or S.Y.
**Author contributions**
S.Y. conceived the experiments. P.K., Z.Y., C.W. and S.Y. carried out the STM experiments and experimental data analysis. G.L., Y.Q. and C.F. were responsible for sample growth. C.W. and S.Y. wrote the manuscript with input from all authors.
**Notes**
The authors declare no competing financial interest.

## ACKNOWLEDGEMENTS

We acknowledge J. Liu and N. Hao for fruitful discussions. S.Y. acknowledges the financial support from the National Key R&D Program of China (Grant No. 2020YFA0309602), National

Science Foundation of China (Grant No. 11874042) and the start-up funding from ShanghaiTech University. C.W. acknowledges the support from National Natural Science Foundation of China (Grant No. 12004250) and the Shanghai Sailing Program (Grant No. 20YF1430700). Y.Q. acknowledges the financial support from the Science and Technology Commission of Shanghai Municipality (Grant No. 19JC1413900). G.L. and C.F. are financially supported by the European Research Council (ERC Advanced Grant No. 742068 TOPMAT). G.L. and C.F. also acknowledge the funding by the DFG through SFB 1143 (project ID. 247310070) and the Würzburg-Dresden Cluster of Excellence on Complexity and Topology in Quantum Matter ct.qmat (EXC2147, project ID. 39085490) and via DFG project HE 3543/35-1. G.L. is supported by the Foundation of the President of Ningbo Institute of Materials Technology and Engineering, Chinese Academy of Sciences.

**Figures:**

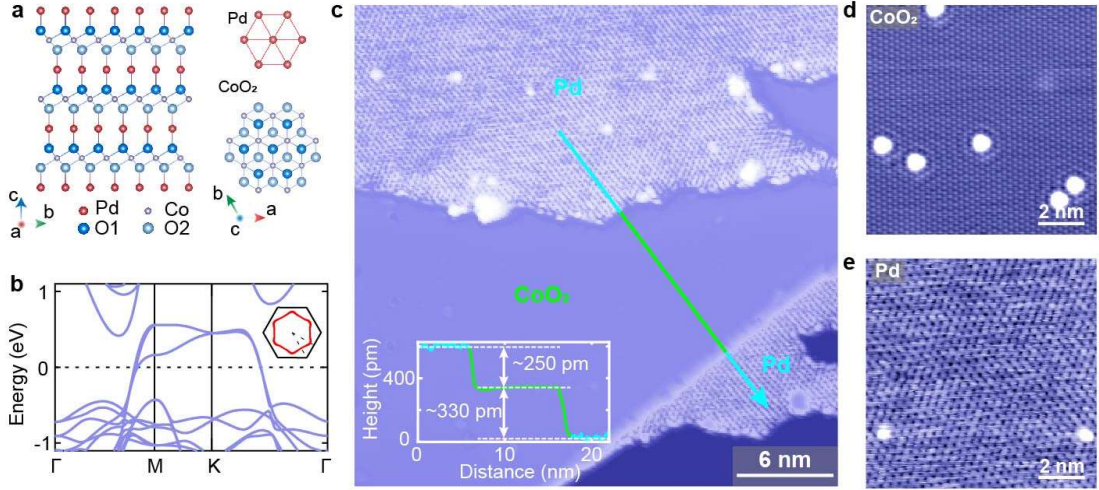

**Figure 1.** (a) Crystal structure of PdCoO$_2$. (Left: side view along the "a" axis. Right: top view along the "c" axis on the Pd-terminated surface (top) and the CoO$_2$-terminated surface (bottom)). (b) The previously calculated band structure for the bulk PdCoO$_2$ adapted with permission from Ref. 27, APS. The inset illustrates the bulk Fermi surface and the Brillouin zone. (c) Constant-current topography taken on a large area of the cleaved PdCoO$_2$ sample, showing several atomic steps. The line profile shows the step height along the colored line. ($V_s$ = 500 mV, $I$ = 20 pA). (d, e) Constant-current STM topographies taken on the CoO$_2$- and Pd-terminated surfaces, respectively (c: $V_s$ = −200 mV, $I$ = 500 pA; d: $V_s$ = 500 mV, $I$ = 20 pA).

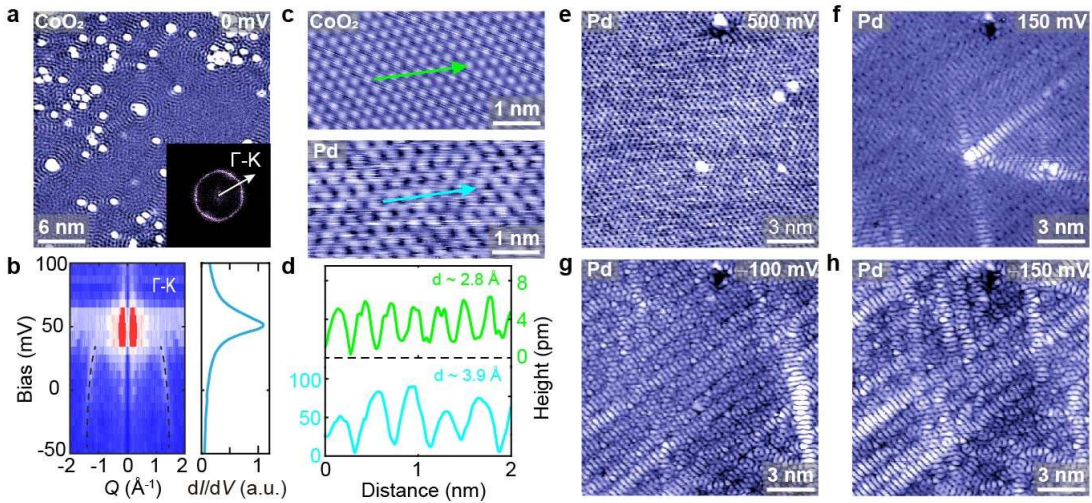

**Figure 2.** (a) d$I$/d$V$ map on the CoO$_2$-terminated surface at the bias voltage of 0 mV. The inset is QPI pattern by performing FT to the d$I$/d$V$ map. The white arrow in the inset marks the Γ-K direction. (b) Linecut along the Γ-K direction in the QPI pattern as a function of energy (left). The typical

d$I$/d$V$ spectrum taken on the CoO$_2$-terminated surface (right). (c) Constant-current STM topographies taken on the CoO$_2$-terminated surface (upper: $V_s$ = −200 mV, $I$ = 500 pA) and the Pd-terminated surface (lower: $V_s$ = 500 mV, $I$ = 20 pA;). (d) Line profiles along the green and the blue arrows in (c). (e)-(h) Constant-current topographies taken on the Pd-terminated surface with (e) +500, (f) +150, (g) −100 and (h) −150 mV bias voltages, respectively.

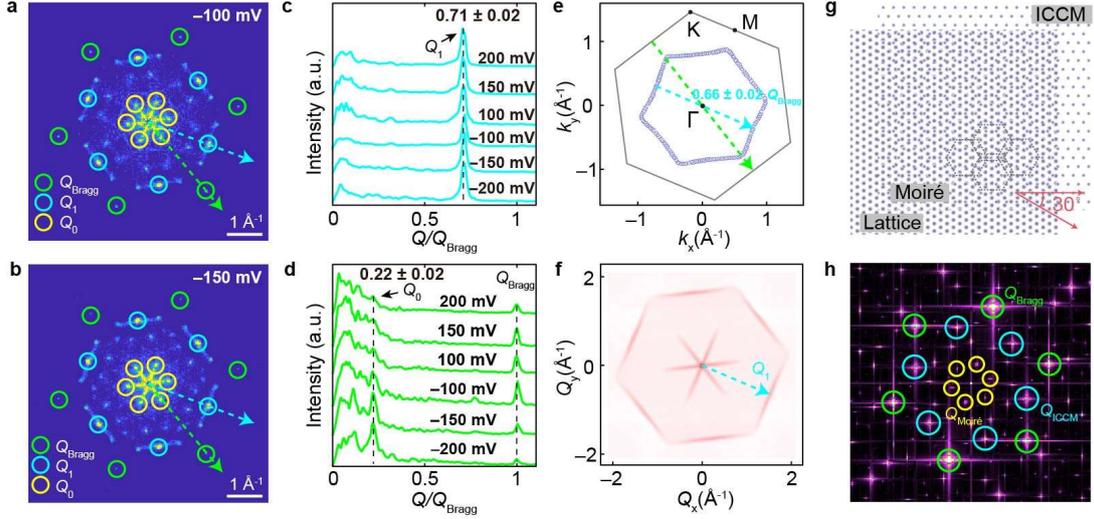

**Figure 3.** (a, b) FT images of the STM topographies shown in Figs. 2 (g) and (h), respectively. The green, blue and yellow circles correspond to the $Q_{Bragg}$, $Q_1$ and $Q_0$ wavevectors, respectively (More FT images are shown in Fig. S4). (c, d) Line profiles taken along the directions shown by the blue (c) and the green (d) arrows in (a, b) for the FT images of the STM topographies taken with various bias voltages. (e) Contour of the Brillouin zone (the grey hexagon) and the measured Fermi surface (the dotted hexagon) contributed by the Pd-terminated surface of PdCoO$_2$. The data are extracted from Ref. 17. The blue arrow indicates the nesting vector of the Fermi surface, and the green arrow shows the Bragg wavevector. (f) Autocorrelation of the Fermi surface in (e). The blue arrow represents the nesting vector in the Fermi surface. (g) Overlap of the lattice and the ICCM pattern with a lattice angle of 30˚, forming an interference pattern. (h) FT image of the overlapping region in panel (g). The green, blue and yellow circles indicate the $Q_{Bragg}$, $Q_{ICCM}$ and $Q_{Moiré}$ wavevectors, respectively.

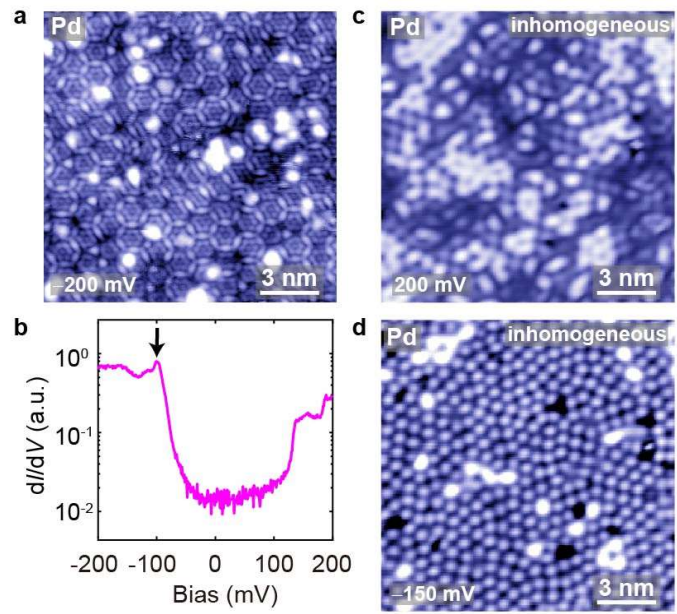

**Figure 4.** (a) Constant-current STM topography taken on the Pd-terminated surface ($V_s = -200$ mV, $I = 20$ pA). (b) The typical d$I$/d$V$ spectrum taken on the Pd-terminated surface ($V_s = -200$ mV, $I = 0.5$ nA). (c, d) Constant-current STM topographies taken on the inhomogeneous Pd surface with positive and negative bias voltages, respectively (c: $V_s = 200$ mV, $I = 300$ pA; d: $V_s = -150$ mV, $I = 100$ pA).